# An Investigation into the Correlations between Students' Written Responses to *Lecture-Tutorial* Questions and their Understandings of Key Astrophysics Concepts

**Abstract**

This paper reports on an investigation into the correlations between students' understandings of introductory astronomy concepts and the correctness and coherency of their written responses to targeted *Lecture-Tutorial* questions. We assessed the correctness and coherency of responses from 454 college-level, general education, introductory astronomy students enrolled in courses taught in the spring of 2010, 2011, and 2012. We hypothesized that students who consistently provided responses with high levels of correctness and coherency would outperform students who did not on multiple measures of astronomy content knowledge.  We compared students' correctness and coherency scores to their normalized gains on the Light and Spectroscopy Concept Inventory (LSCI) and to their scores on closely-related exam items. Our analysis revealed that no significant correlations exist between students' correctness and coherency scores and their LSCI gain scores or exam item scores.  However, the participant group in our study did achieve high scores on the LSCI and on closely-related exam items. We hypothesized that these differences are due to the discussions that take place between students which suggests that instructors who teach with active engagement instructional strategies should focus their implementation on ensuring that their students fully engage in the richest possible discourse.

**Introduction**
Experts working in discipline-based education research (DBER) have successfully created several different active engagement instructional strategies, spanning multiple STEM disciplines, that lead to well-documented improvements in students' discipline knowledge and skills (Singer, Nielsen, and Schweingruber, 2012; Freeman, *et al*., 2014; Springer, Stanne, and Donovan, 1999).  The development of these instructional strategies is often informed by prior research suggesting that students are more likely to achieve significant, long-lasting learning gains when they actively construct their own understandings of a topic through activities that build upon their prior knowledge and intuitions, and explicitly address their conceptual and reasoning difficulties.  These activities purposefully promote peer discourse and allow students to collaboratively develop shared understandings of complex ideas.  This enables students to acquire discipline skills and abilities that they are unlikely to obtain from passive participation in didactic lectures or while working individually (*e.g.*, Clement, Brown, and Zeitsman, 1989; Elby, 2001; Greeno, 1997; McDermott, 1991; Posner, Strike, Hewson, and Gertzog, 1982; Singer, Nielsen, and Schweingruber, 2012; Vygotsky, 1978).

In the astronomy education community, the *Lecture-Tutorials for Introductory Astronomy* (Prather, Slater, Adams, and Brissenden, 2013) are an example of a research-validated active engagement instructional strategy that intellectually engages students on commonly taught topics in introductory, college-level, general education astronomy courses (hereafter Astro 101). The *Lecture-Tutorials* are pencil and paper activities, designed to be completed by pairs of students in 10 to 25 minutes, working together in lecture hall settings after hearing a short lecture on the relevant topic. These collaborative learning activities are driven by carefully sequenced Socratic questions posed in ordinary language, which guide students toward more expert-like understandings of challenging astrophysical topics. Initially, students are asked to examine a novel situation requiring them to apply information just presented in the lecture. The questions that follow are of increasing difficulty and require students to examine a wide variety of scientific representations, graphs, and data sets, which help to continuously engage students in critical discourse with one another and motivate them to evaluate their developing ideas. Several studies have shown that integrating the *Lecture-Tutorials* into the lecture portion of an Astro 101 class can significantly raise students' discipline knowledge and skills beyond what is typically achieved through a lecture alone (Prather *et al.*, 2004; LoPresto and Murrell, 2009; Prather and Brissenden, 2009; Wallace, Prather, and Duncan, 2012).

While the average learning gains achieved by classes using *Lecture-Tutorials* can be significant, the learning gains of individual students can vary greatly (Schlingman, *et al.*, 2012). We suspect that much of this variation is due to the level to which a student intellectually engages while working through a *Lecture-Tutorial*. One potential measure of a student's understanding is the correctness and coherency of the answers and explanations he or she writes in response to the questions in the *Lecture-Tutorial*s. In this study, we analyzed the written responses of 454 students to three to four of the most conceptually challenging questions in the *Lecture-Tutorials*. We used rubrics to analyze and quantify the correctness and coherency of each written response. We then investigated the extent to which the correctness and coherency of these written responses were correlated to other measures of student learning, such as performance on the research-validated Light and Spectroscopy Concept Inventory (LSCI; Bardar *et al.*, 2006; Schlingman, *et al.*, 2012) and on exam items whose content was closely-related to the content of the *Lecture-Tutorial* questions. The LSCI was chosen for this study because many of its items have proven effective at discriminating between students' conceptual understandings (Schlingman et al. 2012). Additionally, its items directly probe the conceptual knowledge and reasoning abilities that the *Lecture-Tutorials* analyzed in this study were designed to improve.

**Research Methods**
*Questions Analyzed*
Our data set includes the written responses of students enrolled in the spring 2010, spring 2011, and spring 2012 versions of an Astro 101 course taught at a large public land grant research university (see below for more details about this course). We carefully examined the written responses of students from these semesters to the following *Lecture-Tutorial* questions: Question 5 from "Binary Stars" (Figure 1), Question 6 from "Newton's Laws and Gravity" (Figure 2) and Question 12 from "Luminosity, Temperature and Size" (Figure 3). See Prather *et al.* (2013) for the complete *Lecture-Tutorials*. We selected these questions for our research because they

1) are conceptually rich, require students to provide detailed explanations of their reasoning, and rely on knowledge gained from other *Lecture-Tutorials* in order to arrive at a correct answer;
2) represent a wide range of question types (from strictly word-based to graphical and pictorial), cover a wide range of topics (including mechanics, stellar properties, and the nature of light and energy), and are designed to develop students' physical intuitions, metacognitive abilities, quantitative reasoning abilities, and graph-reading skills;
3) are closely matched, in terms of their content, to twenty-one difficult, multiple-choice questions used on course exams (which come from the *Lecture-Tutorial* and Ranking Task efficacy studies - Prather *et al.* 2004; Prather and Brissenden, 2009; Hudgins *et al.* 2006), and to the twenty-six multiple-choice questions on the LSCI; and
4) occur toward the end of their respective *Lecture-Tutorials*, which ensures that students have completed the earlier parts of each *Lecture-Tutorial* before attempting to answer the questions in our research.

For the spring 2011 and spring 2012 data sets, we added a fourth question to our analysis: Question 6 from the "Doppler Shift" *Lecture-Tutorial* (Figure 4). This question was added to expand the range of topics analyzed in our study. Note that this question also satisfies the four criteria stated above.

[Figures 1-4 about here]

The instructors also used and vetted these questions in prior semesters to make these questions clear to students and to ensure that students interpret and answer the question they think they are answering so that wrong answers do not arise as a result of reading mistakes.

***Student Population***

The students whose written *Lecture-Tutorial* responses we analyzed for this study were enrolled in an Astro 101 course with an extremely large enrollment: 761 students enrolled in the spring 2010 course, 719 students in spring 2011, and 684 students in spring 2012. These "mega-courses" met only twice a week for one hour and fifteen minutes in the university's largest performing arts center. All three courses were taught by the same instructor using the same curriculum. Additionally, the instructor and instructional staff made every attempt possible to provide equivalent instruction to each of the three courses. The instructional strategies and techniques used by both the instructor and instructional staff were controlled for through weekly meetings on each *Lecture-Tutorial* topic. During these meetings, questioning strategies, common student alternative conceptions, and the best practices to identify, confront and resolve students' reasoning difficulties were discussed to ensure that students received equivalent instruction for all three semesters. Despite the size of these courses and limited instructional resources, we were able to engineer an interactive learning environment in which students achieved learning gains consistent with the top 10% of scores in a recent national study (Prather, Rudolph, and Brissenden, 2011; Prather, Rudolph, Brissenden, and Schlingman, 2009).

To gain access to students' written responses, we asked students to submit their *Lecture-Tutorial* books at the end of each semester in exchange for a small amount of course extra credit (less than 1% of their overall course grade). Note that during the semester students were never asked to submit their *Lecture-Tutorial* activities for grading. 597 students from the spring 2010 semester (78% of the total enrollment), 623 students from the spring 2011 semester (87% of the total enrollment), and 573 students from the spring 2012 semester (84% of the total enrollment) turned in their *Lecture-Tutorials*.

We wanted to examine the correctness and coherency of students' written responses to *Lecture-Tutorial* questions and how this related to students' performance on the LSCI and on closely-related exam items. In order to be included in our data set, a student had to provide answers to
1) all of the *Lecture-Tutorial* questions we analyzed;
2) all of the LSCI items, both pre- and post-instruction; and
3) all of the closely-related exam items.

Note that all students in the class were asked to take the LSCI (both pre- and post-instruction) and were given a nominal amount of participation credit for doing so. Of the students who submitted their *Lecture-Tutorials*, only 26% ($n$=158) of students from the spring 2010 semester, 29% ($n$=180) of students from the spring 2011 semester, and 20% ($n$=116) of students from the spring 2012 semester also met all three criteria listed above. Our data set thus includes a total of 454 students.

*Scoring Rubrics and Calibration Lists*
Our team of researchers analyzed hundreds of students' written responses to the *Lecture-Tutorial* questions until we agreed on the elements that would constitute a well-constructed written response. Using a constant comparative approach (Glaser and Strauss 1967), our researchers independently coded samples of students' written responses based on their factual accuracy and cohesiveness. We then compared these codes, negotiating discrepancies as they arose. By this process, we were able to identify patterns within students' responses that we considered to be of the highest quality and expert-like. We found that the highest quality responses posses two attributes. The first attribute we coded for was the scientific accuracy (which we refer to as "correctness") of each piece of reasoning present within the response. The second attribute coded was the inter-connectedness of pieces of reasoning as part of a larger explanation (which we refer to as "coherency.") The most correct and coherent written responses drew evidence and reasoning from multiple sources, built upon topics discussed in prior *Lecture-Tutorials*, and contained multiple connections between different pieces of reasoning.

To quantify students' written responses to *Lecture-Tutorial* questions, we designed a set of detailed rubrics used to score the correctness and coherency of students' answers and explanations. These scores constitute our data on students' written responses (see Figures 5 and 6 for the correctness and coherency rubrics for Question 6 of "Newton's Laws and Gravity".) These scoring rubrics identify what is meant by a "correct piece of reasoning," and a "correct," "complete," "coherent," or "interconnected" explanation. Each rubric also contains a question-specific criteria list, which provides additional information to assist researchers in consistently coding students' responses. In addition to these rubrics, we created calibration lists of student responses for each question (see Figure 7 for the calibration list for Question 6 of "Newton's Laws and Gravity"). These calibration lists are composed of real student responses that exemplify the variety of written responses provided by students. To calibrate our researchers, each researcher used both rubrics to independently code the student responses included in the calibration lists. The researchers then compared these codes and resolved any discrepancies. This process allowed us to identify and control sources of researcher bias before coding the written responses. The final versions of our rubrics were used to evaluate both the correctness and coherency of students' written responses on a 0 to 3 scale for each of the *Lecture-Tutorial* questions we investigated. Note that some student responses can have identical correctness and/or coherency codes, even though they vary from one another. Also note that students' written responses could achieve a 3-level score for "coherency" without providing a "correct" answer.

[Figures 5-7 about here]

To assess the inter-rater reliability of our researchers, we examined the frequency with which our researchers' codes to written *Lecture-Tutorial* responses agreed or disagreed with one another. Table 1 contains the values of Cohen's $\kappa$ (Cohen, 1960) for three pairs of researchers. Using the $\kappa$ value interpretation guidelines provided by Landis & Koch (1977), our results indicate an almost perfect agreement level of inter-rater reliability between the researchers on our correctness and coherency rubrics for the spring 2011 and 2012 samples. This suggests that our correctness and coherency rubrics can be used reliably amongst multiple researchers to systematically code students' written *Lecture-Tutorial* responses.

[Table 1 about here]

**Results and Discussion**
*Distribution of Combined Correctness and Coherency Scores*
Every student in our study received both a correctness and a coherency score for each question we investigated. Once scored, we summed the correctness and coherency scores across all questions in our study for each student. Thus, we used one number for each student (the combined score of all of his or her correctness and coherency scores) to characterize the overall quality of his or her written responses across all questions analyzed. For the rest of this paper, we will refer to this number as the "combined correctness and coherency score." It is worth noting that we also performed analyses of students' individual scores for correctness or coherency separately, but found no more significant or unique findings than what is reported below.

Figures 8-10 show the distributions of students' combined correctness and coherency scores for all three semesters. Note that a different range of combined correctness and coherency scores is possible for the spring 2010 sample than for the spring 2011 and 2012 samples. This difference is due to the fact that we added a question (Question 6

from the "Doppler Shift" *Lecture-Tutorial*) to our analysis of the spring 2011 and 2012 data.

[Figures 8-10 about here]

Almost all students in our study fall within two standard deviations of each sample's mean. Few students ($n \leq 5$) are represented in the points with the lowest or highest combined correctness and coherency scores. These distributions provide evidence that almost *no* students consistently provide incorrect and incoherent answers. Likewise, almost *no* students consistently provide correct and coherent written responses at the 3-level. The fact that so few students reside at the extremes of the distributions may suppress our ability to detect potential correlations between students' combined correctness and coherency scores and other measures of student learning.

### *Correlations between written Lecture-Tutorial responses and other measures of student achievement*

Our research was focused on determining the extent to which students' levels of understanding of key ideas from their introductory astronomy course were related to the level of writing they engaged in while completing their *Lecture-Tutorial* activities. After scoring students' written responses, we scored students' responses to the LSCI and to the closely-related exam items. These multiple-choice items were scored as either 1 (correct) or 0 (incorrect). We then summed these scores to produce a total score for each student on the LSCI (both pre- and post-instruction) and on the exam items. We then calculated each student's normalized gain score on the LSCI. Next, we analyzed the correlations between students' scores for correctness and coherency with their level of understanding of astronomy concepts as measured by their normalized gain scores and exam scores.

We used Pearson's coefficient of determination ($R^2$) to determine whether or not a correlation exists between students' combined correctness and coherency scores and their LSCI normalized gain and exam item scores. An $R^2=0$ indicated that no correlation was found, whereas an $R^2=1$ indicated that a perfect correlation was found. These $R^2$ values are illustrated in Table 2 .

[Table 2 about here]

These results suggest that students who achieved higher combined correctness and coherency scores did not outperform students with lower combined correctness and coherency scores. We observed no significant correlation between students' correctness and coherency scores and their normalized gain scores or their scores to related exam items. These results taken alone would suggest that *Lecture-Tutorials* had no affect on students' learning of astrophysical concepts. However, as we will discuss below, we believe the analysis of the data given thus far provides an incomplete picture from which to gauge the success or failure of the *Lecture-Tutorials* with regard to improving student astronomy discipline knowledge.

### *Did completing the Lecture-Tutorials actually help students learn?*

Since the distributions of students' normalized gain scores and exam item scores cover approximately the same range of values across all combined correctness and coherency scores, it is tempting to conclude that the *Lecture-Tutorials* have no effect on students' understandings of astronomy. However, students' normalized gain scores and exam scores provide compelling evidence that significant amounts of learning took place each semester. Using the gain scores calculated for each student, we averaged these scores and found the average normalized gain is $<g>=0.46$ for the spring 2010 sample, $<g>=0.48$ for the spring 2011 sample, and $<g>=0.47$ for the spring 2012 sample. These values are higher than what is achieved by the vast majority of Astro 101 classes for these topics and place these classes in the top 10% of courses in the nation (Prather *et al.* 2009).

Students in our sample also did well on the challenging exam items we investigated: 74% correct in spring 2010, 70% correct in spring 2011 and 69% correct in spring 2012. Note that these items were chosen not only for their topical relationship to the *Lecture-Tutorial* questions we investigated but also because these questions are conceptually challenging and target known conceptual and reasoning difficulties. The high averages on these items are well above what is typically achieved after lecture alone, and are consistent with post-*Lecture-Tutorial* averages from prior studies (Prather *et al.* 2004; Prather and Brissenden, 2009).

These high average normalized gain scores and high average exam item scores provide strong evidence to support the claim that these students made significant improvements in their knowledge of the assessed astrophysics topics. In addition to investigating students' achievement on the LSCI and closely-related exam items, we also investigated the degree to which these students constitute a representative sample of Astro 101 students. We will next discuss our investigation into whether the students' who participated in our study started the course with a greater initial understanding of the astronomy topics we investigated and if they had greater discipline knowledge after instruction than the average Astro 101 student.

***Are the student participants in our study representative of their classes?***
One way to test the representativeness of our sample is to look at the initial and final knowledge states of students, both inside and outside of our sample, as measured by the LSCI and course final exam scores. We compared the LSCI and final exam data of students who met our study criteria (hereafter referred to as the participant group) to a group of students (hereafter referred to as the comparison group) who did not complete all of the *Lecture-Tutorial* questions included in our study but did complete the LSCI (pre- and post-instruction) and the final exam. We found that the average pre-instruction LSCI scores and standard deviations for the participant group (22% ± 11% in the spring 2010, 24% ± 9% in the spring 2011, and 24% ± 9% in the spring 2012) are very close to the average pre-instruction scores and standard deviations of the comparison group (24% ± 10% in the spring 2010, 22% ± 9% in the spring 2011, and 25% ± 9% in the spring 2012). These averages for the participant and comparison groups are also close to the 24% ± 2% average pre-instruction LSCI scores for the 69 different classes participating in a recent national study of Astro 101 courses (Prather *et al.* 2009). These data provide strong evidence that the participating students in our sample do not begin the semester with deeper conceptual understandings than their peers or compared to the national population of Astro 101 students.

However, the data suggest that students in the participant and comparison groups ended the class with different levels of conceptual understanding. Figure 11 shows the distribution of final exams scores of the participant and comparison groups. Figure 12 shows the distribution of final course grades of the participant and comparison groups. The participant group had average final exam grades that were 6-8% higher than those of the comparison group. A greater percentage of students in the participant group also achieved grades of "A" and "B" for the course overall than the students in the comparison group. It is tempting to infer that the students who participated in our study were *not* representative of the students in their classes. This impression is bolstered when one recalls that, in order to be included in our sample, a student had to answer all of the *Lecture-Tutorial* questions we analyzed. However, based on the pre-instructional LSCI scores, we assert that the students from each semester started on equal footing with regard to their understanding of difficult astrophysics concepts. So, if the students in the participant and comparison groups start off with the same level of understandings, how can we account for their difference in post-instruction understanding? As we discuss further below, we hypothesize that students in the participant group represent students who engaged more consistently and participated more frequently in the in-class peer discussions promoted by the *Lecture-Tutorials*.

[Figures 11 and 12 about here]

**Comments and Conclusions**
This paper reports on an investigation into the relationship between students' understandings of introductory astronomy concepts and the correctness and coherency of their written responses to targeted *Lecture-Tutorial* questions. We assessed the correctness and coherency of responses from 454 college-level general education introductory astronomy students enrolled in courses taught at the same state university by the same instructor during the spring semesters of 2010, 2011, and 2012. We hypothesized that students who consistently provided answers and explanations with high levels of correctness and coherency would outperform their peers who did not on multiple measures of astronomy content knowledge.

Many of the outcomes of this study were quite surprising to our research group. Prior studies conducted by the authors and others had documented the success of *Lecture-Tutorials* at helping students develop more expert-like understandings of astrophysics than what is typically achieved from traditional lecture alone. We were surprised at the near zero correlation between students' written *Lecture-Tutorial* responses and their achievement on the LSCI and related exam items. We believe this lack of correlation may be partially explained by the surprisingly small number of students in our participant group who consistently gave high-level written responses to the *Lecture-Tutorial* questions (resulting in a data set utterly lacking in high combined correctness and coherency scores). This result lead us to further investigate the level of understanding of these students relative to their peers.

It would be erroneous to conclude that students in our sample did not learn any of the astronomy content addressed by their *Lecture-Tutorials*. Our data show exactly the opposite. Our participant groups had average normalized gain scores on the LSCI that places them among the top 10% of scores in a recent national study of Astro 101 courses, and their average correctness on the closely-related exam items used in these courses were well above what is typically achieved from lecture alone, and consistent with post-*Lecture-Tutorial* averages from prior studies (Prather *et al.* 2004; Prather and Brissenden, 2009).

The pre-instruction LSCI scores of our participant group revealed that these students began class with conceptual understandings and reasoning abilities that were, on average, the same as their peers and equivalent to Astro 101 students from other courses taught around the country, suggesting that our participant group did not begin the course with superior abilities or content mastery of astrophysics. However, we observed that more students in our participant group achieved grades of "A" and "B" on their final exams and their final course grades than students in our comparison group.

One possible explanation for this difference is that the *Lecture-Tutorials* can promote high levels of intellectual engagement and collaboration between students. While working collaboratively on their *Lecture-Tutorials*, students are asked to repeatedly engage in sense-making discussions and defend the reasoning behind their answers to the *Lecture-Tutorial* questions. We suspect that students in the participant group were among the most regular and earnest participants in these in-class activities and discussions, since to be included in the participant group a student must have written down answers to all of the analyzed *Lecture-Tutorial* questions, completed the LSCI pre- and post-instruction, and answered all of the closely related exam items. The written responses we analyzed may not reflect the transformations in conceptual understandings and reasoning abilities that occur during students' discourse. To further investigate this hypothesis, future studies may directly measure the effect that peer-to-peer discourse has on improving students' discipline knowledge and skills.

5) Stars that are very close together will often orbit around one another, and, occasionally, their orbits are aligned in such a way that one star will pass directly in front of the other as seen from Earth. These stars are often referred to as eclipsing binary stars. Which of the two times (1 or 2) labeled below most likely indicates the time when the Sun-like (G-spectral type) star was passing **in front** of the A-spectral type star from the previous questions (*circle 1 or 2*)? Explain your reasoning.

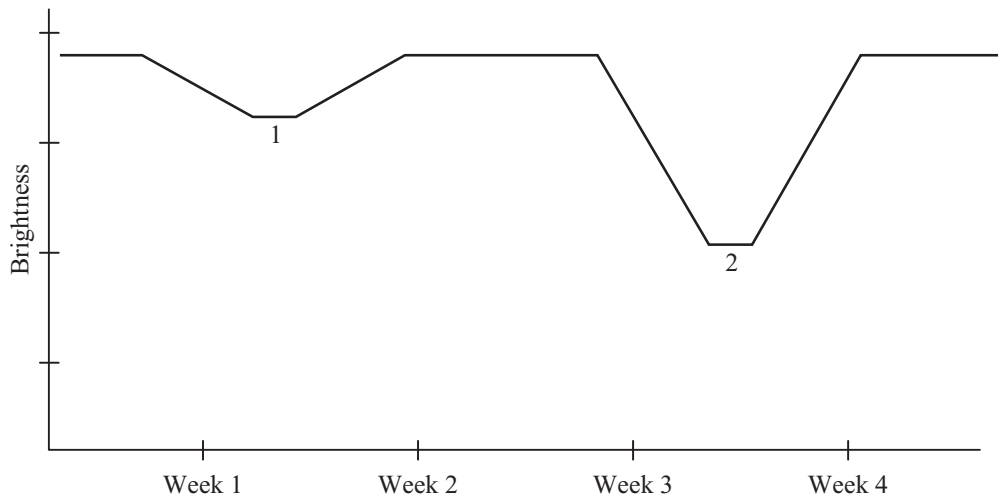

**Figure 1.** Question 5 from the "Binary Stars" *Lecture-Tutorial* (Prather *et al*., 2013).

In the picture below, a spaceprobe traveling from Earth to Mars is shown at the halfway point between the two (not to scale).

6) Where would the spaceprobe experience the strongest net (or total) gravitational force exerted on it by Earth and Mars? Explain your reasoning.

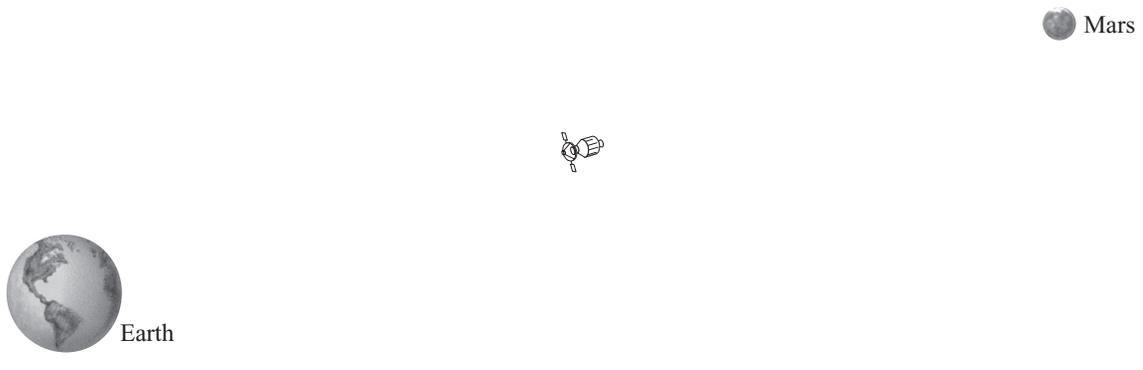

**Figure 2.** Question 6 from the "Newton's Laws and Gravity" *Lecture-Tutorial* (Prather *et al*., 2013).

The graph below plots the luminosity of a star on the vertical axis against the star's surface temperature on the horizontal axis. This type of graph is called an H-R diagram. Use the H-R diagram below and the relationship between a star's luminosity, temperature, and size (as described on the previous page) to answer the following questions concerning the stars labeled U-Y.

5) Based on the information presented in the H-R diagram, which star is larger, X or Y? Explain your reasoning.

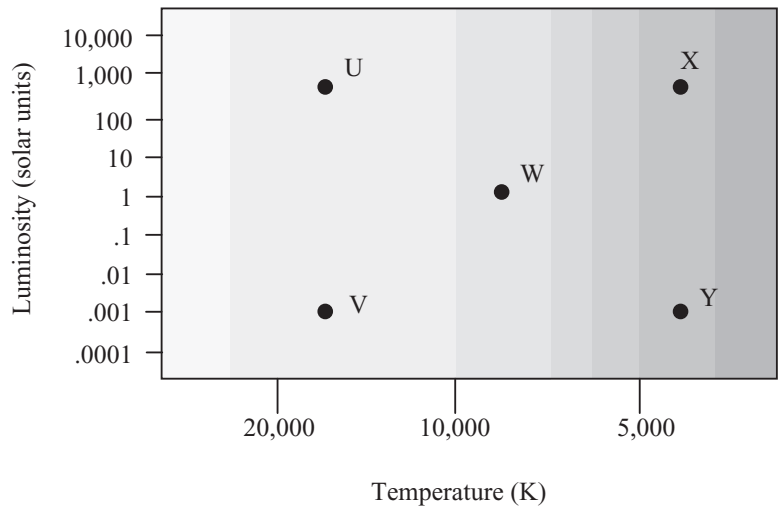

**Figure 3.** Question 12 from the "Luminosity, Temperature, and Size" *Lecture-Tutorial* (Prather *et al.*, 2013).

For the three absorption line spectra shown below (A, B, and C), one of the spectra corresponds to a star that is not moving relative to you, one of the spectra is from a start that is moving toward you, and one of the spectra is from a star that is moving away from you.

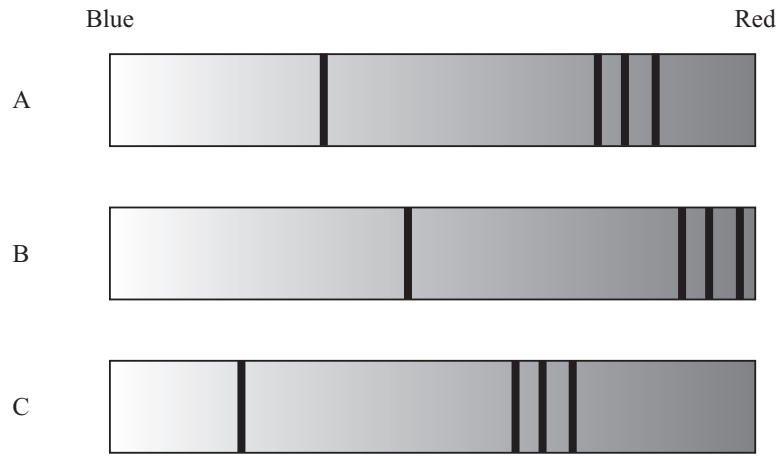

6) Which of the three spectra above corresponds with the star moving toward you? Explain your reasoning.

**Figure 4.** Question 6 from the "Doppler Shift" *Lecture-Tutorial* (Prather *et al*., 2013).

| Lecture-Tutorial Research Project – Correctness Rubric |
|---|
| **Scoring Term definitions** |
| A **student response** is anything recorded in a student's Lecture-Tutorial. |
| A **correct piece of reasoning** is any factual statement that is relevant to the topic. |
| An **incorrect piece of reasoning** is any idea that contains physically incorrect, ambiguous, or irrelevant information. |
| A **correct explanation** contains at least one correct piece of reasoning and no incorrect pieces of reasoning. |
| An **incorrect explanation** contains at least one incorrect piece of reasoning. |
| A **complete explanation** contains at least three correct pieces of reasoning. |
| An **incomplete explanation** contains one or two correct pieces of reasoning. |
| A correct & complete explanation contains at least three correct pieces of reasoning that support the student's answer. Explanations where only one correct piece of reasoning is present are correct but incomplete, because they do not contain multiple pieces of reasoning. If the explanation does not support the student's answer, it's an incorrect explanation. Below is a description of the different levels awarded for different levels of correctness. |

**Rubric Score Levels**

| Score 0: | Score 1: | Score 2: | Score 3: |
|---|---|---|---|
| The student response contains:<br><br>• An incorrect piece of reasoning.<br><br>Their explanation is incorrect and incomplete. | The student response contains:<br><br>• At least one correct piece of reasoning.<br><br>Their explanation is incomplete. It can contain other correct but irrelevant pieces of reasoning. | The student response contains:<br><br>• Two different correct pieces of reasoning.<br><br>Their explanation is correct and incomplete. | The student response contains:<br><br>• At least three different correct pieces of reasoning.<br><br>Their explanation is correct and complete. |

**Newton's Law Question #6 Specific Criteria – Correctness**
- Writing "Earth" by itself is an indication that you place the satellite on or very close to Earth to experience the strongest net gravitational force.
- "Closer to Earth" does not mean "On Earth" or "Closest to Earth". It is an incorrect piece of reasoning.
- Describing Net Force at zero is not a correct piece of reasoning.
- If there is anything written in the explanation area do not grade the LT as blank.

*Figure 5:* The correctness rubric for "Newton's Laws and Gravity" Question 6.

| *Lecture-Tutorial* Research Project – Coherency Rubric |
|---|
| **Scoring Term definitions** |
| A **student response** is anything recorded in a student's *Lecture-Tutorial*. |
| A **complete piece of reasoning** is any idea that is clear, concise, and relevant to the topic. |
| An **incomplete piece of reasoning** is any idea about the topic that is ambiguous, irrelevant, or fragmented. |
| A **comparison** contrasts two different pieces of reasoning. |
| A **connection** strings together two different pieces of reasoning. |
| A **coherent explanation** contains a single, complete piece of reasoning or multiple, complete pieces of reasoning. |
| An **incoherent explanation** contains an incomplete piece of reasoning. |
| An **unconnected explanation** contains a single, complete piece of reasoning but does not connect it to another, complete piece of reasoning. |
| A **singularly connected explanation** makes a connection between two different pieces of reasoning. |
| An **interconnected explanation** contains multiple, complete pieces of reasoning and also compares or connects several different complete pieces of reasoning together. |
| A coherent & interconnected explanation compares or connects several pieces of reasoning together to provide support for an answer. Responses where a single piece of reasoning is present are coherent but unconnected explanations, because they do not connect multiple pieces of information together to support an argument. Below is a description of the different levels awarded for different levels of coherency. |

| Rubric Score Levels | | | |
|---|---|---|---|
| **Score 0**: | **Score 1**: | **Score 2**: | **Score 3**: |
| The student response contains:<br><br>• An incomplete piece of reasoning.<br><br>Their explanation is incoherent and unconnected. | The student response contains:<br><br>• A single, complete piece of reasoning.<br><br>Their explanation is coherent but unconnected. | The student response contains:<br><br>• A complete piece of reasoning connected to a different, complete piece of reasoning.<br><br>Their explanation is coherent and singularly connected. | The student response contains:<br><br>• Multiple connections between at least three different, complete pieces of reasoning<br><br>Their explanation is coherent and interconnected. |

**Newton's Law Question #6 Specific Criteria – Coherency**
- Distance and location are not different pieces of reasoning.
- Complete pieces of reasoning that are connected to incomplete pieces of reasoning cannot be given a score of 2 because two different, complete pieces of reasoning connected together are necessary to make a 2-level response.

Figure 6: The coherency rubric for "Newton's Laws and Gravity" Question 6.

| Lecture-Tutorial Research Project -- Calibration List for "Newton's Laws and Gravity" Question 6 ||
|---|---|
| Student Responses:<br><br>(A) The largest net gravitational force would be closest to the Earth b/c the force on the satellite from the Earth is stronger than the force on the satellite by Mars, so when you subtract the forces it will be larger.<br>(B) Closer to Mars, because Earth has more mass and the closer the space probe gets to Mars, the stronger the force from Mars.<br>(C) The Earth b/c it has a greater mass.<br>(D) Closest to the biggest mass would = greatest force.<br>(E) In the middle, because the forces are the same.<br>(F) A little closer to Mars so force becomes equal.<br>(G) Slightly to the right, because the gravitational pull would be intersecting and would be between the two.<br>(H) It is in the Earth's gravitational pull. Making it have the strongest force.<br>(I) Closer to Earth cuz it has a stronger force.<br>(J) In the middle because they are within equidistance of each other therefore the exertion of gravity is being pulled at the exact rate & force.<br>(K) The middle because if two things are pulling on each other they have the same gravitational pull.<br>(L) Closer to Mars with the same pull.<br>(M) When the force from both planets is equal.<br>(N) Closest to Earth. The strongest gravity between Earth and the weakest between Mars.<br>(O) When it is closest to Earth because Earths pull is at its greatest and Mars is at its weakest.<br>(P) (point labeled "strongest" drawn next to Earth) The space probe would experience the strongest next force. | (Q) Because you need to take distance/mass into account.<br>(R) (point labeled "A" drawn next to Earth) A, because the mass is larger. (S) The forces would be equal and it would cause it to have a strong pull both ways.<br>(T) When the gravitational pull from Earth pulls the space probe closer.<br>(U) Location in question 4.<br>(V) Where it is.<br>(W) Does not matter because the Net force never changes.<br>(X) Closest to the biggest mass.<br>(Y) Somewhere closer to Earth w/o cancelling out Mars' pull.<br>(Z) (point labeled "Strongest force" drawn next to Earth) The closer the space probe gets to a planet, the stronger the net force is.<br>(AA) When it's closest to Earth or Mars.<br>(AB) If the space probe was not in the middle of the planets it would have a strong force because you add the forces together.<br>(AC) It would be on each.<br>(AD) Earth > Mars<br>(AE) (point labeled "Strongest force from Earth" drawn next to Earth) Whichever planet has the largest mass would create the largest pull if it was close.<br>(AF) At the point right before it hits Earth. Net force & acceleration are always in the same direction.<br>(AG) The forces would be equal and it would cause it to have a strong pull both ways. |

**Figure 7.** The calibration list for "Newton's Laws and Gravity" Question 6.

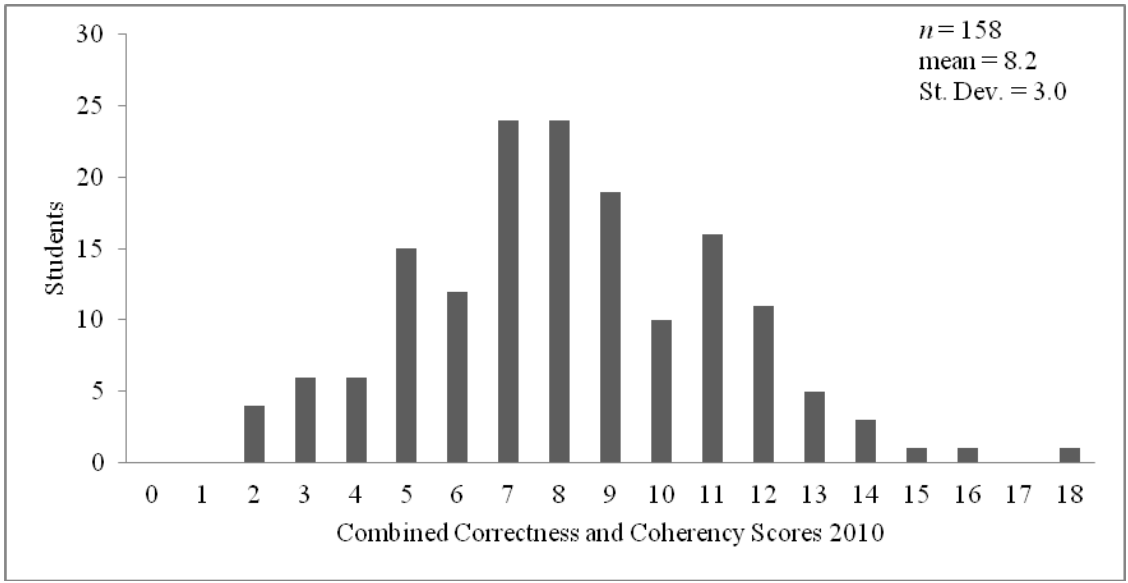

**Figure 8.** The distribution of combined correctness and coherency *Lecture-Tutorial* scores for the spring 2010 sample.

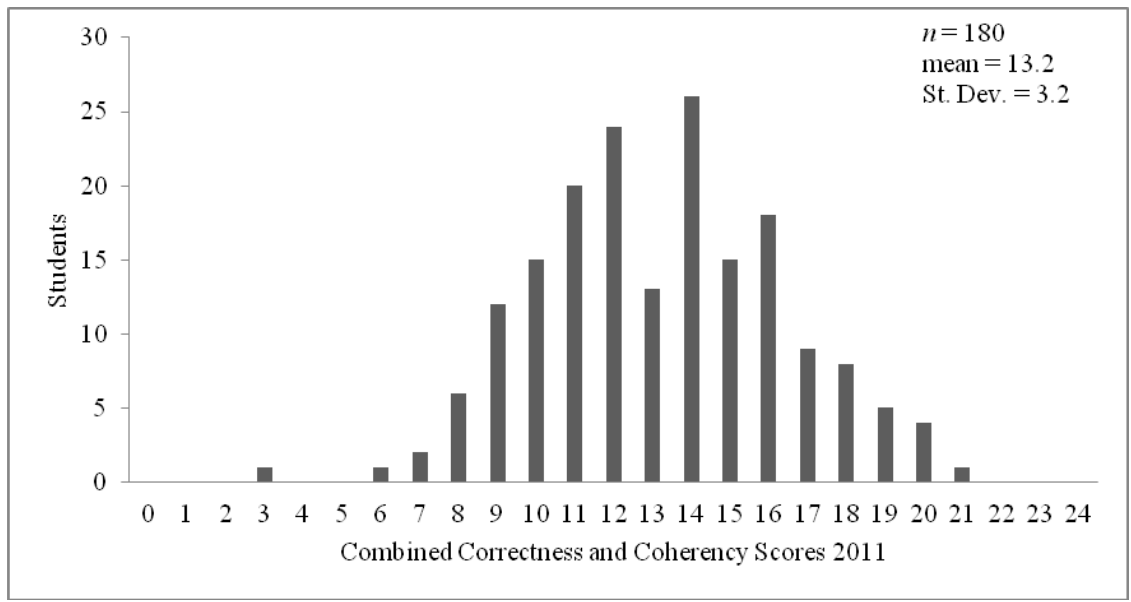

**Figure 9.** The distribution of combined correctness and coherency *Lecture-Tutorial* scores for the spring 2011 sample.

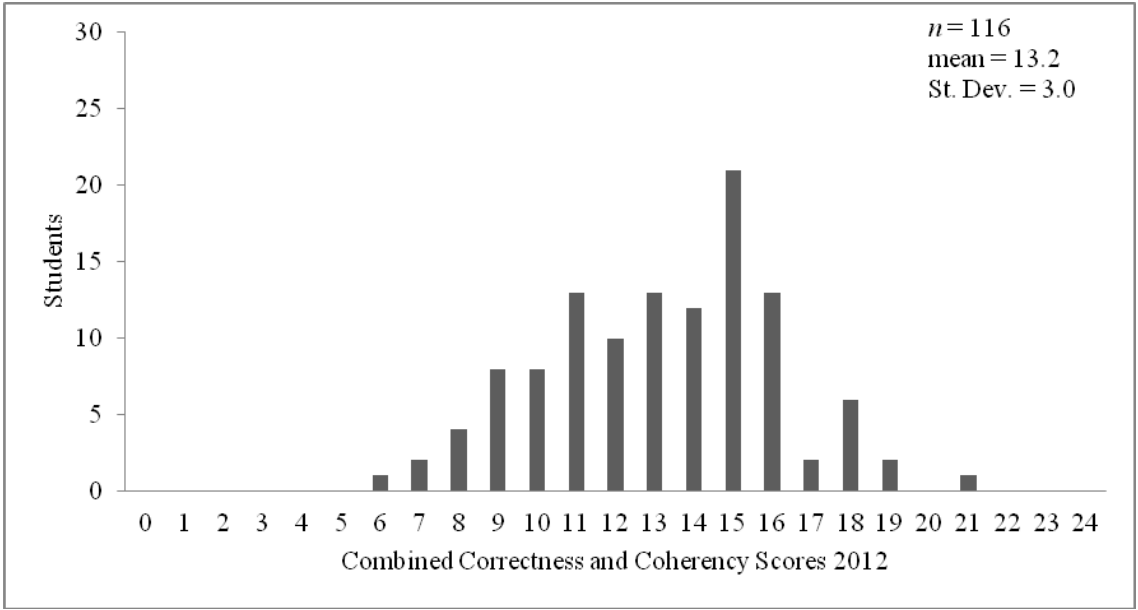

**Figure 10.** The distribution of combined correctness and coherency *Lecture-Tutorial* scores for the spring 2012 sample.

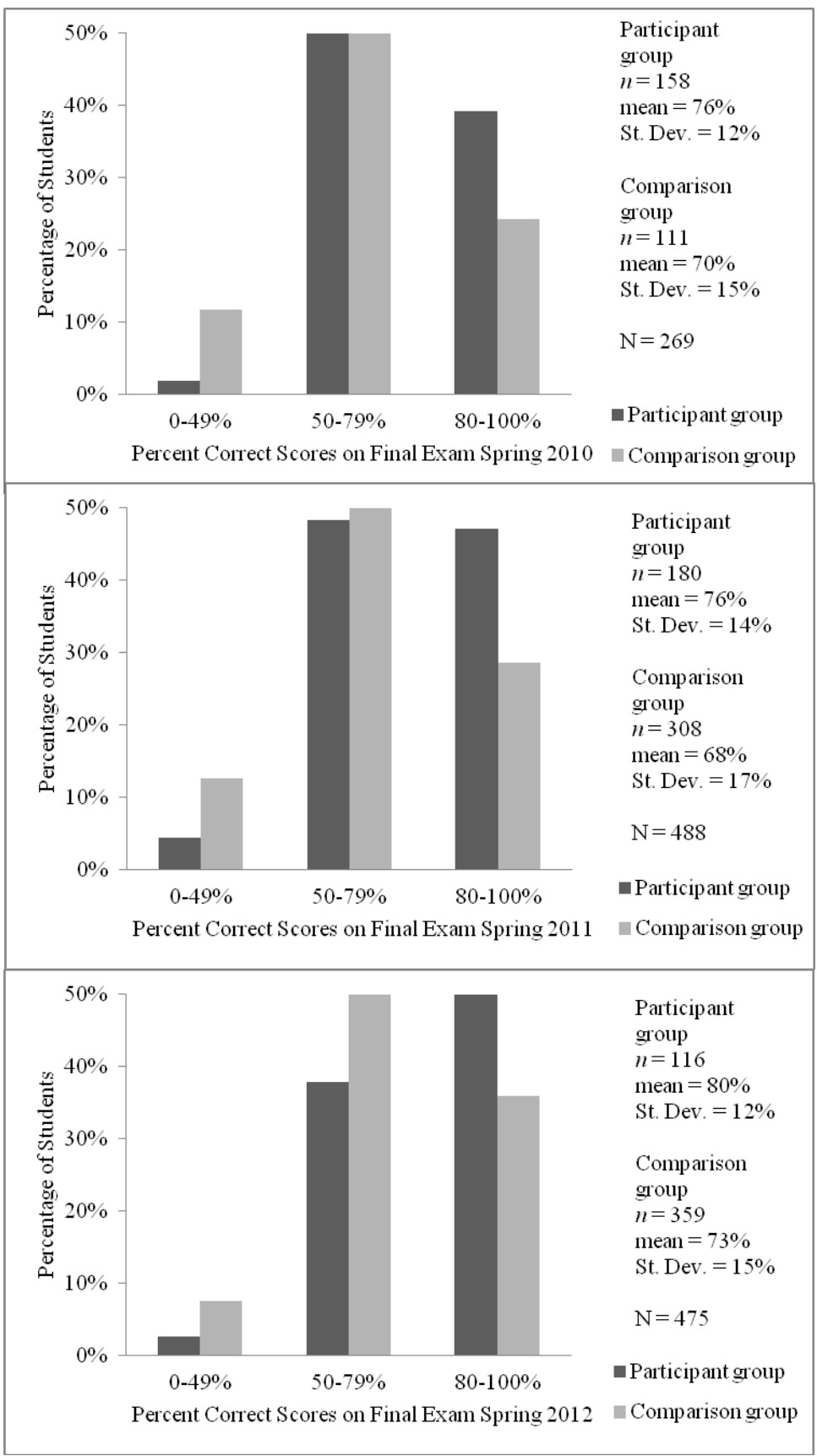

**Figure 11.** Final exam scores distributions for the spring 2010, 2011 and 2012 semesters, respectively.

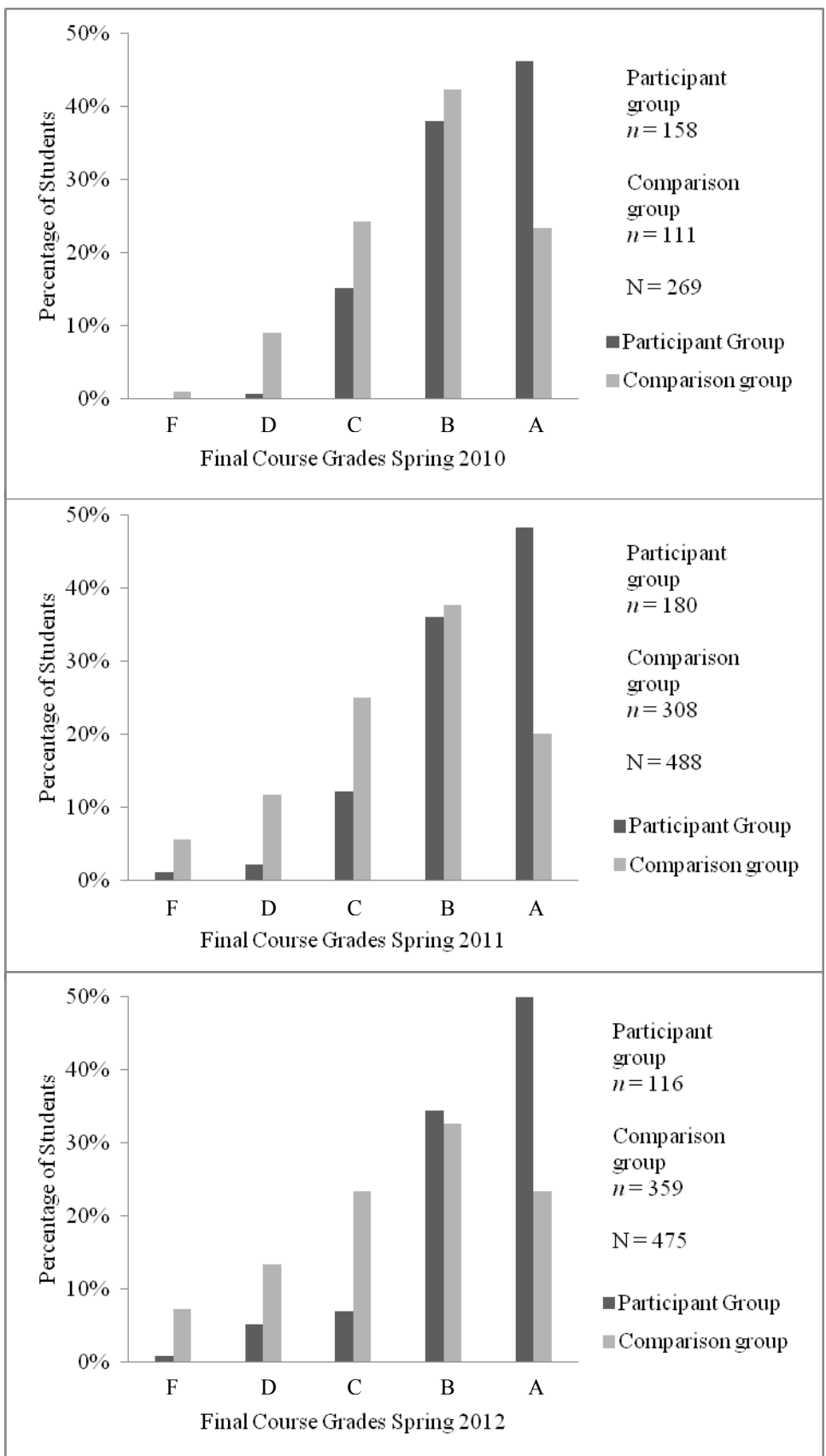

**Figure 12.** Final grade distributions for the spring 2010, 2011 and 2012 semesters, respectively.

| Table 1. Cohen's $\kappa$ values and standard errors (SE) between three pairs of researchers (Spring 2011 and 2012) | | | | | | | |
|---|---|---|---|---|---|---|---|
| | | Researcher 2 | | Researcher 3 | | Researcher 4 | |
| | | $\kappa$ | SE | $\kappa$ | SE | $\kappa$ | SE |
| Researcher 1 | Correctness | 0.839 | 0.020 | 0.909 | 0.020 | 0.927 | 0.020 |
| | Coherency | 0.826 | 0.020 | 0.911 | 0.020 | 0.929 | 0.020 |

| Table 2. Analysis of correlation between Combined Correctness and Coherency Scores, LSCI normalized gain scores, and percent correct on related exam items (Spring 2010, 2011 and 2012) | | | |
|---|---|---|---|
| | Combined Correctness and Coherency Scores | | |
| | Spring 2010 ($n$=158) | Spring 2011 ($n$=180) | Spring 2012 ($n$=116) |
| LSCI normalized gain scores | 0.012 | 0.043 | 0.047 |
| Percent correct on related exam items | 0.017 | 0.037 | 0.000 |